\documentclass{ws-ijmpd}

\begin{document}

\markboth{Owen Pavel Fern\'{a}ndez Piedra}
{Gravitino perturbations in Schwarzschild black holes}

%
\catchline{}{}{}{}{}
%

\title{GRAVITINO PERTURBATIONS IN SCHWARZSCHILD BLACK HOLES}

\author{OWEN PAVEL FERNANDEZ PIEDRA}

\address{$^{1}$ Departamento de F\'{i}sica y Qu\'{i}mica, Universidad de Cienfuegos, Carretera a Rodas, Cuatro Caminos, s/n. Cienfuegos, Cuba\\
opavel@ucf.edu.cu}

\address{$^{2}$ Instituto de F\'{\i}sica, Universidade de S\~ao Paulo,
  CP 66318,
05315-970, S\~ao Paulo, Brazil\\
jeferson@fma.if.usp.br}

\maketitle


\begin{abstract}
We consider the time evolution of massless gravitino perturbations
in Schwarzschild black holes, and show that as in the case of fields
of other values of spin, the evolution comes in three stages, after
an initial outburst as a first stage, we observe the damped
oscillations characteristic of the quasinormal ringing stage,
followed by long time tails. Using the sixth order WKB method and
Prony fitting of time domain data we determine the quasinormal
frequencies. There is a good correspondence between the results
obtained by the above two methods, and we obtain a considerable
improvement with respect to the previously obtained third order WKB
results. We also show that the response of a black hole depends crucially on the
spin class of the perturbing field: the quality factor becomes a decreasing function of the spin for boson perturbations
, whereas the opposite situation appears for fermion ones.
\end{abstract}


\maketitle

\section{Introduction}

The study of the evolution of small perturbations in black hole
backgrounds is a very interesting subject. Actually, we know that
this evolution, at intermediate times, is  dominated by damped
single frequency oscillations. These characteristic oscillations
have been termed quasinormal modes and the associated frequencies
quasinormal frequencies, and depend only on the parameters
characterizing the black hole, as it mass, electric charge and
angular momentum \cite{nollert,kokkotas1,berti1}. In this sense we can say that black
holes have a characteristic sound, resembling for example the
familiar sound produced bay the ringing of a bell or the strum of a
guitar.

Once one realizes the importance of black holes in fundamental
physics, one can grasp the meaning and significance of their
characteristic vibrations \cite{nollert,kokkotas1,berti1}. Black holes have
been called the hydrogen atom of general relativity, perfect
comparison because like the hydrogen atom in quantum mechanics,a
black hole, as a solution of the Einstein's field equations has all
the general relativistic properties embodied in it, but still is
simple enough to be a model for starting a complete understanding of
all the physics that go with Einstein's gravitation theory.

There are different contexts in which the study of quasinormal modes
of black holes appears to be motivated:  the estimation of
astrophysical black hole parameters \cite{berti1,berti2}
, the study of the stability of these solutions under
small perturbations, semiclassical ways to quantize the black hole
area and the estimation of thermalization timescales in connection
with the ADS/CFT correspondence, in which a large static black hole
in asymptotically AdS spacetime corresponds to a thermal state in a
Conformal Field Theory ( CFT ) at its boundary, and the decay of the
test field in the black hole spacetime corresponds to the decay of
the perturbed state in the CFT. In this sense, we can determine the
timescale for the return to thermal equilibrium studying its
dynamics in AdS spactime, and then translating it onto the CFT,
using the AdS/CFT conjecture \cite{horowitz} \cite{Miranda}.

Since the discovery of quasinormal oscillations of black holes, many
studies have been done on quasinormal modes of various spin fields and a
considerable variety of analytical, semi-analytical and numerical
methods have been developed to determine them \cite{chandrasekar1,chandrasekar2,leaver1,leaver2,abdalla,gpp,carlemoskono,abdwangetc,shutz-will,iyer-will,konoplya1,konoplya2,zhidenkothesis}.

An interesting problem to consider is the study of quasinormal modes
of spin $s=3/2$ fields, called Rarita-Schwinger fields. In a previous
paper Shu and Shen \cite{shu} determined the quasinormal modes of this fields in
Schwarzschild spacetime using the third
order WKB method developed by Iyer and Will \cite{iyer-will}.

In this paper we investigate for the first time the complete time
evolution of gravitino perturbations in the Schwarzschild background
and determine with better accuracy the quasinormal frequencies by
fitting time domain data by a superposition of damping exponents,
using the Prony method. We also extend previous results by Shu and
Zheng beyond the third order WKB, using the sixth order approach
developed by Konoplya \cite{konoplya1,konoplya2}. The results obtained by this
two approaches match very well, ensuring the validity of the
determined quasinormal frequencies. We also studied the complete time evolution of scalar, electromagnetic, gravitational and Dirac perturbations
in the mentioned spacetime, and compute numerically some quasinormal frequencies by the two methods mentioned above.

The structure of this paper is as follows. In Section II we use the Newmann-Penrose formalism to obtain the equations for gravitino test perturbations in Schwarzschild spacetime, and use separation of variables to obtain decoupled equations for the radial and angular variables involved. The radial equation is written in the familiar way useful to study the quasinormal spectrum of the perturbations, obtaining an analytical expression for the effective potential related to it. Section III is devoted to determine the time domain evolution and to compute some quasinormal frequencies of gravitino perturbations, and to compare them with similar calculations for other boson and fermion fields. The last section contains our conclusions and presents some lines for future research on this subject.

\section{Rarita-Schwinger equation in Schwarzschild spacetime}

In supergravity theory the gravitational field is coupled to the
massless spin 3/2 Rarita-Schwinger field, that acts as a source of
torsion and curvature. if we denote by $\psi_{AB\dot{C}}$ the spin
3/2 field, then the supergravity fields equations are invariant
under supergravity transformations
$\psi_{AB\dot{C}}\rightarrow\psi_{AB\dot{C}}+\nabla_{B\dot{C}}\alpha_{A}$,
where $\alpha_{A}$ is an arbitrary spinor field.

When the
Rarita-Schwinger field vanishes, the supergravity equations becomes
Einstein's vacuum equations. Therefore, since the torsion and the
curvature produced by the Rarita-Schwinger field depend
quadratically on this field, then in the linear approximation about
a solution with $\psi_{AB\dot{C}}=0$, the supergravity field
equations reduces to the Rarita-Schwinger field equations together
with the vacuum Einstein's field equations \cite{aichelburg,guven}.

The Rarita-Schwinger
equations for the field $\psi_{AB\dot{C}}$ in a curved background
can be written as
\begin{equation}\label{}
    H_{ABC}=H_{\left(ABC\right)},\ \ \ \ H_{A\dot{B}\dot{C}}=0\label{RSeq}
\end{equation}
where we have defined $H^{A}_{\ \ BC}\equiv\nabla_{(B}^{\ \ \
\dot{D}}\psi^{A}_{\ \ C)\dot{D}}$ and $H^{A}_{\ \
\dot{B}\dot{C}}\equiv\nabla^{D}_{\ (\dot{B}}\psi^{A}_{\ \
|D|\dot{C})}$ (the parenthesis denotes symmetrization on the indices
enclosed and the indices between bars are excluded from the
symmetrization). It is possible to show that in an algebraically special vacuum spacetime, the contraction of
$H_{AB\dot{C}}$ defined above, with a multiple principal spinor of
the conformal curvature satisfies a decoupled equation \cite{torres1}.

Using the
Newmann-Penrose notation, we can show that in a frame such that
$\Psi_{2}$ is the only nonvanishing component of the Weyl spinor,
the Rarita-Schwinger reduces to only two equations for the
components $H_{000}\equiv\Omega_{\frac{3}{2}}$ and
$H_{111}\equiv\Omega_{-\frac{3}{2}}$, that can be solved by
separation of variables in all the type D vacuum backgrounds. In terms of the differential operators $\widehat{A}= \left(D-2\epsilon+\epsilon^{*}-3\rho-\rho^{*}\right)(\widetilde{\Delta}-3\gamma
    +\mu)$, $\widehat{B}= \left(\delta-2\beta-\alpha^{*}-3\tau+\pi^{*}\right)\left(\delta^{*}
    -3\alpha+\pi\right)$, $\widehat{C}=(\widetilde{\Delta}+2\gamma-\gamma^{*}+3\mu+\mu^{*})\left(D+3\epsilon-\rho\right)$ and $\widehat{D}=\left(\delta^{*}+2\alpha+\beta^{*}+3\pi
    -\tau^{*}\right)\left(\delta+3\beta-\tau\right)$ that contains the usual directional derivatives of the Newmann-Penrose formalism, we can put the Rarita-Schwinger equations in the above mentioned frame as
\begin{equation}\label{}
    \left(\widehat{A}-\widehat{B}-\Psi_{2}\right)\Omega_{\frac{3}{2}}=0\label{RSeqnp1}
\end{equation}
\begin{equation}\label{}
    \ \ \left(\widehat{C}-\widehat{D}-\Psi_{2}\right)\Omega_{-\frac{3}{2}}=0\label{RSeqnp2}
\end{equation}
We are interested in the specific case of Schwarzschild background,
where the metric can be expressed as
\begin{equation}\label{}
   ds^{2}=\frac{\triangle}{r^{2}}dt^{2}-\frac{r^{2}}{\triangle}dr^{2}-r^{2}\left(d\theta^{2}+\sin^{2}\theta
     d\phi^{2}\right)\label{metric}
\end{equation}
with $\triangle=r^{2}-rr_{H}$ and $r_{H}=2M$ defines the location of
the event horizon. The separable solution of (\ref{RSeqnp1}) and
(\ref{RSeqnp2}) are given by
\begin{equation}\label{}
    \Omega_{\frac{3}{2}}=R_{\frac{3}{2}}(r)S_{\frac{3}{2}}(\theta)e^{i(\omega t+m\phi)}\label{separation1}
\end{equation}
\begin{equation}\label{}
    \Omega_{-\frac{3}{2}}=-\frac{1}{2\sqrt{2}r^{3}}R_{-\frac{3}{2}}(r)S_{-\frac{3}{2}}(\theta)e^{i(\omega t+m\phi)}\label{separation2}
\end{equation}
where $\omega$ is the frequency, $m$ is a half integer, and the
functions $R(r)$ and $S(\theta)$ satisfy the ordinary differential
equations
\begin{eqnarray}\label{}\nonumber
\left(\triangle\mathfrak{D}_{-\frac{1}{2}}\mathfrak{D}_{0}^{\dagger}-4i\omega
r\right)\triangle^{\frac{3}{2}}R_{\frac{3}{2}}(r)=\lambda\triangle^{\frac{3}{2}}R_{\frac{3}{2}}(r)
 \\
    \left(\triangle\mathfrak{D}^{\dag}_{-\frac{1}{2}}\mathfrak{D}_{0}+4i\omega
r\right)R_{-\frac{3}{2}}(r)=\lambda R_{-\frac{3}{2}}(r)\label{Reqs}
\end{eqnarray}

\begin{eqnarray}\label{}\nonumber
\mathfrak{L}^{\dag}_{-\frac{1}{2}}\mathfrak{L}_{-\frac{3}{2}}S_{\frac{3}{2}}(\theta)=-\lambda
S_{\frac{3}{2}}(\theta)
 \\
    \mathfrak{L}_{-\frac{1}{2}}\mathfrak{L}^{\dag}_{-\frac{3}{2}}S_{-\frac{3}{2}}(\theta)=-\lambda
S_{-\frac{3}{2}}(\theta)\label{Seqs}
\end{eqnarray}
where $\lambda$ is a separation constant that takes the values
$(\ell+3)(\ell+1)$ with $\ell=2,3,4...$. In the above equations we
have used the operators defined by
\begin{eqnarray}\label{}
\mathfrak{D}_{n}\equiv\partial_{r}+\frac{i\omega
r^{2}}{\triangle}+n\frac{d}{dr}\left(\ln\triangle\right),\ \ \
\mathfrak{D}_{n}^{\dag}\equiv\partial_{r}-\frac{i\omega
r^{2}}{\triangle}+n\frac{d}{dr}\left(\ln\triangle\right)
\\
    \mathfrak{L}_{n}\equiv\partial_{\theta}+m\csc\theta+n\cot\theta,
    \ \ \
    \mathfrak{L}^{\dag}_{n}\equiv\partial_{\theta}-m\csc\theta+n\cot\theta
    \ \ \ \
\end{eqnarray}
Now introducing the tortoise coordinate $r^{*}$ defined by
$dr^{*}=(r^{2}/\triangle)dr$ and making use of the variable
$Y(r)=r^{-2}R_{-\frac{3}{2}}(r)$ we can put the second equation in
(\ref{Reqs}) in the standard form
\begin{equation}\label{}
    \Lambda^{2}Y+P\Lambda_{+}Y-QY=0 \label{eqcasifinal}
\end{equation}
where $\Lambda_{\pm}\equiv\frac{d}{dr^{*}}\pm i\omega$,
$\Lambda^{2}\equiv\Lambda_{+}\Lambda_{-}=\Lambda_{-}\Lambda_{+}=\frac{d^{2}}{dr_{*}^{2}}+\omega^{2}$
and
\begin{equation}\label{}
    P(r)=\frac{3}{2r^{4}}\left(4\triangle+rr_{H}-2r^{2}\right)
\end{equation}
\begin{equation}\label{}
    Q(r)=\frac{\triangle}{r^{5}}\left((\ell+3)(\ell+1)r+r_{H}\right)
\end{equation}
As explained in Reference \cite{torres2}, it suffices to consider
the equation for $R_{-\frac{3}{2}}$ only, and an entirely similar
reduction can be made with the equation for $R_{\frac{3}{2}}$.
Finally. It is possible to reduce (\ref{eqcasifinal}) to a,
Schrodinger-type equation in the form
\begin{equation}\label{}
    \frac{d^{2}}{dr_{*}^{2}}Z(r)+\left[\omega ^{2}-V(r)\right]Z(r)=0 \label{finaleq}
\end{equation}
where the effective potential for the Rarita-Schwinger field is
given by
\begin{equation}\label{}
    V(r)=Q(r)-\frac{dT(r)}{dr_{*}}\label{potential}
\end{equation}
with the function $T(r)$ defined by $T(r)=-\frac{1}{F(r)}(\ell+3)(\ell+1)\sqrt{(\ell+3)(\ell+1)+1}\,+\,\frac{d}{dr_{*}}(\ln
F(r))$, and
$F(r)=\frac{r^{6}}{\triangle^{3/2}}Q(r)$.

\begin{figure}[htb!] 
           \includegraphics[width=11cm]{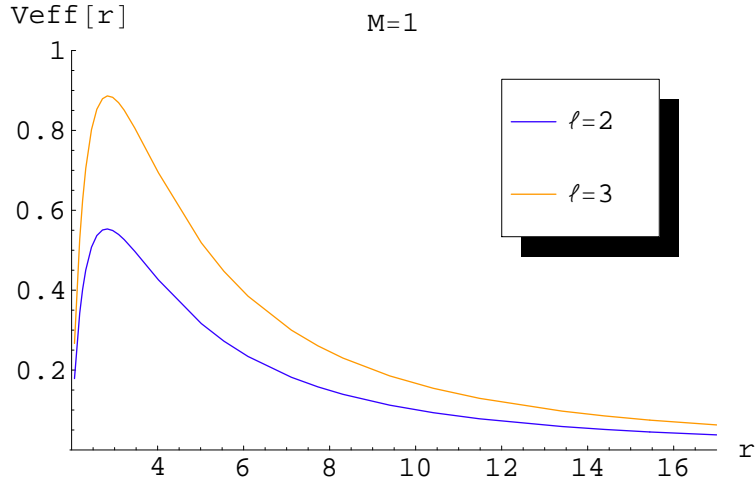}\\
           \caption{Effective potential of gravitino modes for Schwarzschild black hole with $\ell=3$ (top) and $\ell=4$(bottom) for $r_{H}=2$.}
           \label{potencial}
       \end{figure}
In Figure (\ref{potencial}) we show the effective potential for gravitino perturbations
with two different multipole numbers
$\ell$, for a Schwarzschild background with $r_{H}=2$. As we can see, $V$ has the form of a definite positive
potential barrier, i.e, it is a well behaved function that goes to
zero at spatial infinity and gets a maximum value near the event
horizon.

\section{Time domain evolution of perturbations and quasinormal frequencies}
\begin{figure}[htb!] 
           \includegraphics[width=11cm]{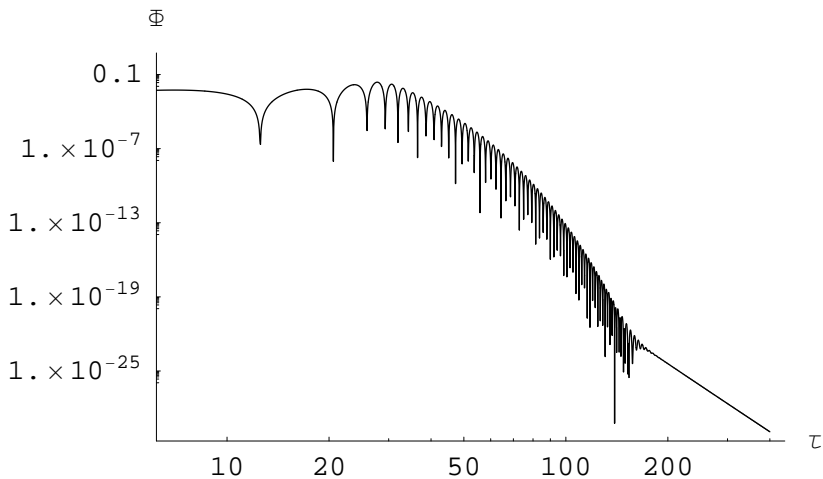}\\
           \caption{\it Logaritmic plot of the time-domain evolution of $\ell=2$ gravitino perturbations of gravitino modes for Schwarzschild black hole at $r=11r_{H}$. In the figure the time is measured in units of the horizon radious: $\tau\equiv t/r_{H}$. }
           \label{perfil1}
       \end{figure}

The most direct approach to study the gravitino perturbations is to
solve numerically the evolution equation associated with
(\ref{finaleq}), that is
\begin{equation}\label{}
    \left(\frac{\partial^{2}}{\partial t^{2}}-\frac{\partial^{2}}{\partial r_{*}^{2}}+V(r)\right)\Phi=0\label{evoleq}
\end{equation}
where the function $\Phi(t,r)$ results from the factorization of
$\Omega_{-\frac{3}{2}}$ in  (\ref{separation2}) as
$\Omega_{-\frac{3}{2}}=\Phi(t,r)S_{-\frac{3}{2}}(\theta)e^{im\phi}$.
In order to integrate numerically the equation (\ref{evoleq}) we use
the technique developed by Gundlach, Price and Pulling \cite{gpp}, and the result can be observed as the time-domain profile
showed in Figure (\ref{perfil1}).

As we can see, the temporal evolution of gravitino perturbation can
be divided in three stages. The first depends on the initial
conditions and the point where we observe the profile. After the
initial outburst at the first stage, we observe the exponential
damping of the perturbations called quasinormal ringing, that can be
split to the superposition of exponentially damping oscillations,
represented by a set of complex frequencies, called quasinormal
frequencies, whose real parts describe the actual frequency of the
oscillation, while the imaginary part is the damping rate.

\begin{figure}[htb!]
\begin{center}
\resizebox{0.45\columnwidth}{!}{\includegraphics*{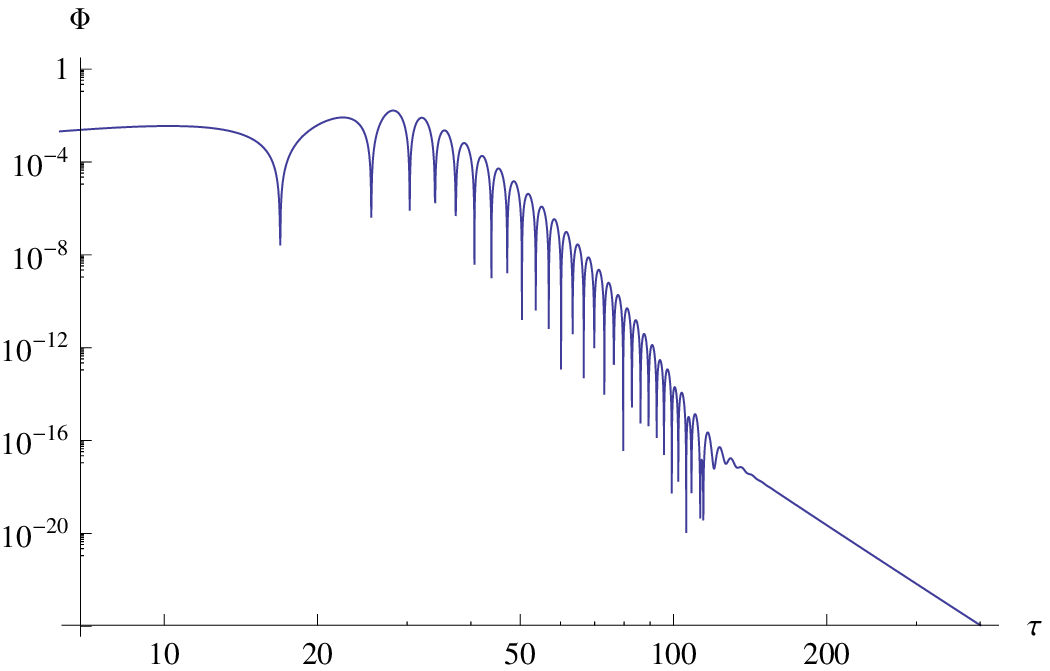}}
\resizebox{0.45\columnwidth}{!}{\includegraphics*{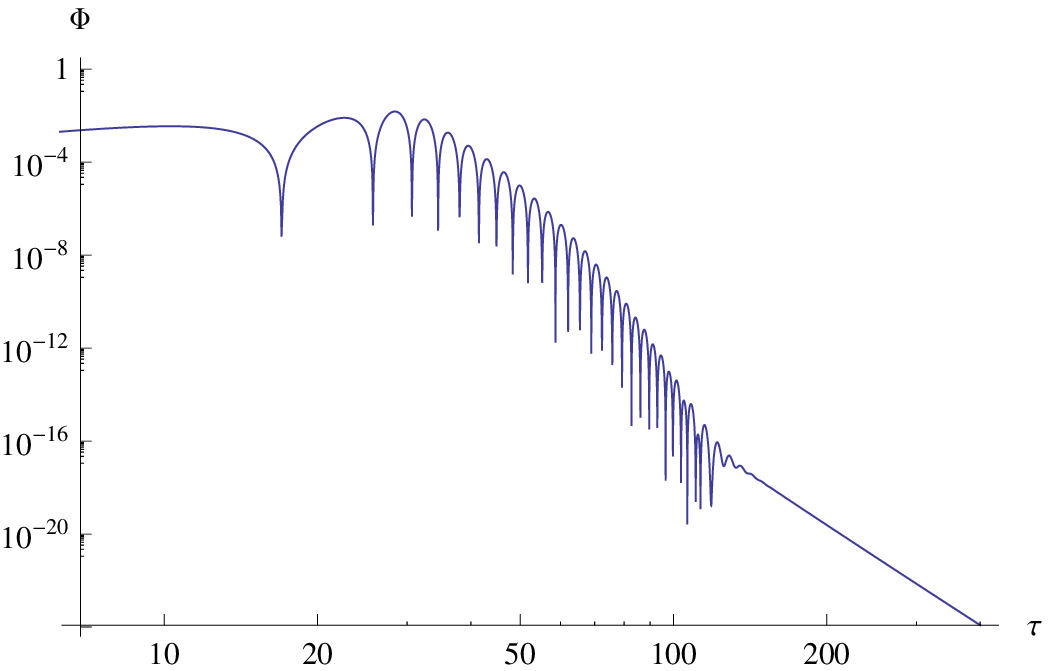}}
\resizebox{0.45\columnwidth}{!}{\includegraphics*{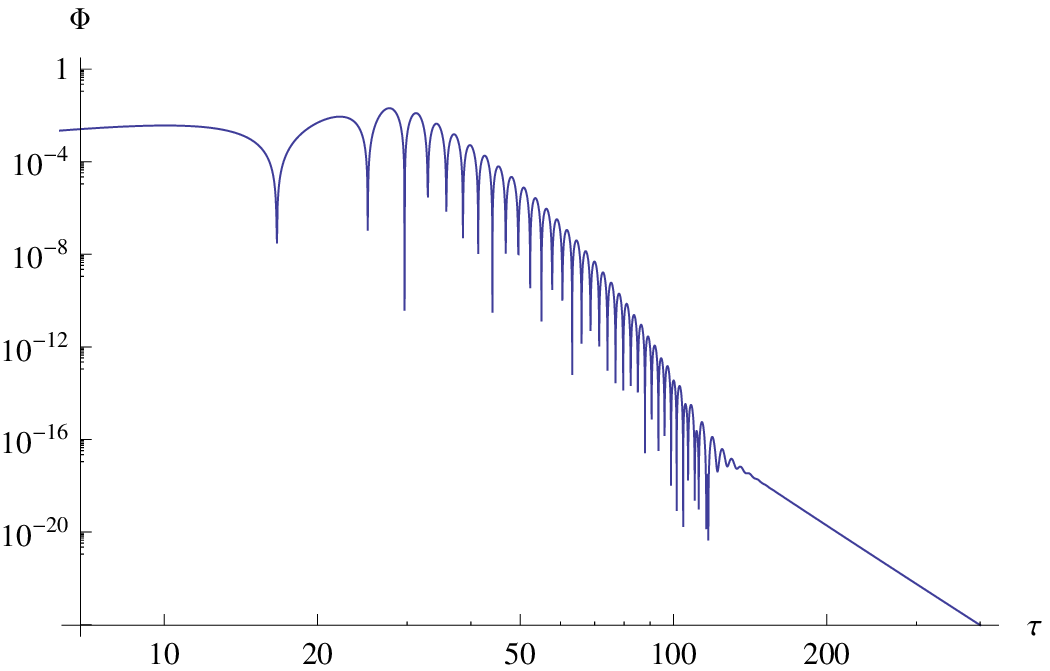}}
\resizebox{0.45\columnwidth}{!}{\includegraphics*{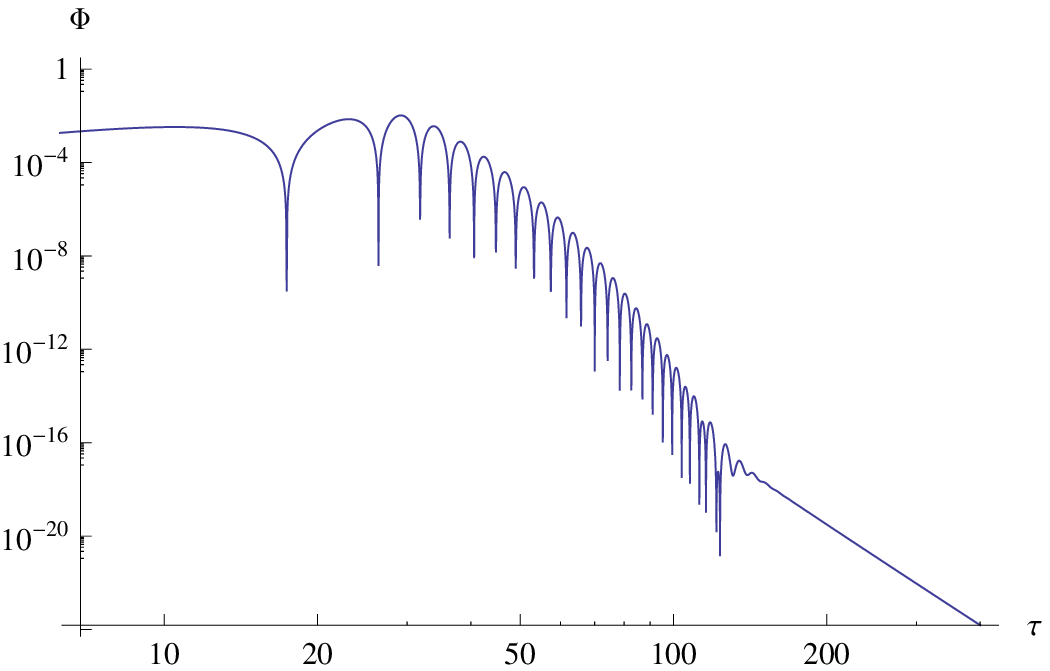}}
\caption{Logaritmic plot of the time-domain evolution of $\ell=2$ scalar (left top), electromagnetic (right top), dirac (left bottom) and gravitational (right bottom) perturbations in Schwarzschild black hole at $r=11r_{H}$.  In the figure the time is measured in units of the horizon radious: $\tau\equiv t/r_{H}$.}
\label{perfilspines}
\end{center}
\end{figure}

The quasinormal modes are solutions of the wave equation (\ref{finaleq})
with the specific boundary conditions requiring pure out-going waves
at spatial infinity and pure in-coming waves on the event horizon.
Thus no waves come from infinity or the event horizon. As we can see
from the figure, the quasinormal ringing stage is followed by the
so-called tails at asymptotically late times.
Thus, the time evolution of gravitino perturbations in
Schwarzschild spacetime follows the same stages as well known boson and fermion perturbations.
To show this, we also integrated numerically the
perturbation equations corresponding to scalar, electromagnetic, gravitational and Dirac perturbations in this spacetime. The resulting profiles are
observed in Figure (\ref{perfilspines}). Note that in all of the cases considered, the time evolution is similar, then the gravitino is not the exception.
In order to evaluate the quasinormal modes we used two different
methods. The first is a semianalytical method to solve equation
(\ref{finaleq}) with the required boundary conditions, based in a
WKB-type approximation, that can give accurate values of the lowest
( that is longer lived ) quasinormal frequencies, and  was used in
several papers for the determination of quasinormal frequencies in a
variety of systems \cite{WKB6papers}.

The WKB technique was applied
up to first order to finding quasinormal modes for the first time by
Shutz and Will \cite{shutz-will}. Latter this approach was extended
to the third order beyond the eikonal approximation by Iyer and Will
\cite{iyer-will} and to the sixth order by Konoplya
\cite{konoplya1,konoplya2}. We use in our numerical calculation of quasinormal
modes this sixth order WKB expansion. The sixth order WKB expansion
gives a relative error which is about two order less than the third
WKB order, and allows us to determine the quasinormal frequencies
through the formula
\begin{equation}\label{WKB6}
    i\frac{\left(\omega^{2}-V_{0}\right)}{\sqrt{-2V_{0}^{''}}}-\sum_{j=2}^{6}\Pi_{j}=n+\frac{1}{2},
    \ \ \ \ \ n=0,1,2,...
\end{equation}
where \(V_{0}\) is the value of the potential at its maximum as a
function of the tortoise coordinate, and \(V_{0}^{''}\) represents
the second derivative of the potential with respect to the tortoise
coordinate at its peak. The correction terms \(\Pi_{j}\) depend on
the value of the effective potential and its derivatives ( up to the
2i-th order) in the maximum, see \cite{zhidenkothesis} and
references therein.

The second method that we used to find the quasinormal frequencies
was the Prony method \cite{berti3,zhidenkothesis} for fitting the time domain profile data by
superposition of damping exponents in the form
\begin{equation}\label{Prony}
    \psi\left(t\right)=\sum_{k=1}^{p}C_{k}e^{-iw_{k}t}
\end{equation}
Assuming that the quasinormal ringing stage begins at $t=0$ and ends at
$t=Nh$, where $N\geq2p-1$, then the expession (\ref{Prony}) is
satisfied for each value in the time profile data
\begin{equation}\label{Prony2}
    x_{n}\equiv\psi\left(nh\right)=\sum_{k=1}^{p}C_{k}e^{-iw_{k}nh}=\sum_{k=1}^{p}C_{k}z_{k}^{n}
\end{equation}
From the above expression, we can determine, as we know $h$, the
quasinormal frequencies $\omega_{i}$ once we have determined $z_{i}$
as functions of $x_{n}$. The Prony method allows to find the $z_{i}$
as roots of the polinomial function $A(z)$ defined as
\begin{equation}\label{Prony3}
    A\left(z\right)=\prod_{k=1}^{p}\left(z-z_{k}\right)=\sum_{m=0}^{p}\alpha_{m}z^{p-m} \
    \ , \ \ \ \ \ \ \ \ \ \ \ \alpha_{0}=1
\end{equation}
It is possible to show that the unknown coefficients $\alpha_{m}$ of
the polinomial function $A(z)$ satisfy
\begin{equation}\label{Prony5}
    \sum_{m=1}^{p}\alpha_{m}x_{n-m}=-x_{n}
\end{equation}
Solving the $N-p+1\geq p$ linear equations (\ref{Prony5}) for
$\alpha_{m}$ we can determine numerically the roots $z_{a}$ and then
the quasinormal frequencies.

It is important to mention the fact that with the Prony method we
can obtain very accurate results for the quasinormal frequencies,
but the practical application of the method is limited because we
need to know with precision the duration of the quasinormal ringing
epoch. As this stage is not a precisely defined time interval, in
practice, it is difficult to determine when the quasinormal ringing
begins. Therefore, we are able to calculate with high accuracy only
two or, sometimes three dominant frequencies.

In Table \ref{frecuencias gravitino} we show the results for the evaluation of
the first four fundamental quasinormal modes for gravitino
perturbations in Schwarzschild black holes, using the two methods mentioned above.

The parameter entering in the calculation is the black hole mass \(M\), that we take to be
a unit mass.
We also show for comparison the results obtained using a third order
WKB method, that were obtained in reference \cite{shu}. With sixth order WKB approximation we obtain an improve of the results
for the quasinormal frequencies, as we can easily see after comparison with the numerical results obtaining using the Prony fitting of the time domain data.

\begin{figure}[htb!] 
           \includegraphics[width=13cm]{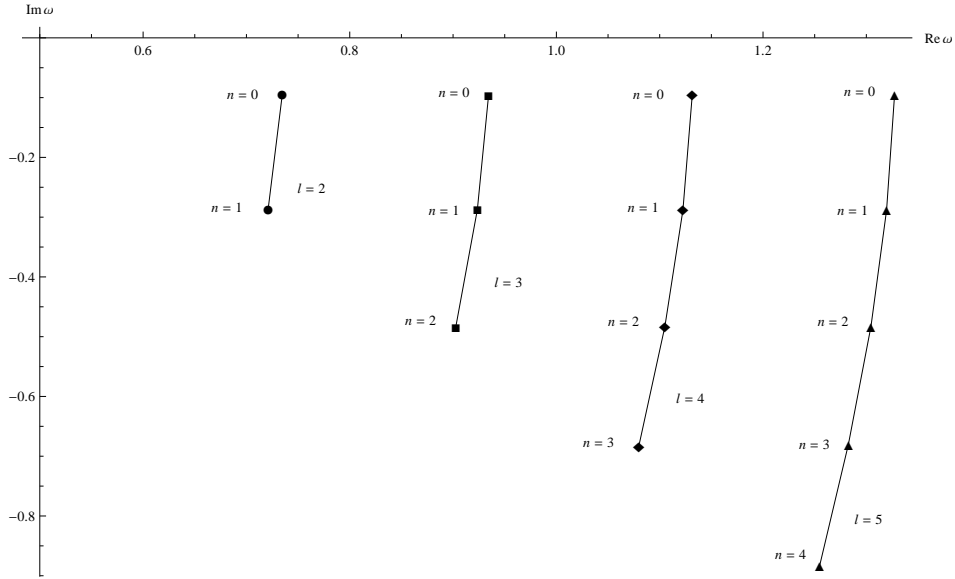}\\
           \caption{\it Gravitino quasinormal modes. }
           \label{qnm}
       \end{figure}

The obtained results are showed in Figure (\ref{qnm}). For a given angular number, higher overtones becomes less oscilatory, i.e, the real part of the quasinormal frequencies decreases whereas the imaginary part increases. In this situation, the quality factor of modes, defined as the ratio $\Upsilon=|Re(\omega)|/|Im(\omega)|$, decreases. On the contrary, modes with higher multipole number $\ell$ and the same overtone number have higher frequencies and the damping shows only little increments, with an effective increase of the quality factors.

To test the accuracy of the Prony method to finding quasinormal frequencies by fitting time domain data, we also
\begin{table}[htb!]
    \tbl{\it Gravitino quasinormal frequencies in Schwarzschild black hole, with
   $M=1$ (third and sixth order WKB approximation and Prony fitting of time domain data).}
  {\begin{tabular}{|c|c|c|c|c|}
      \hline
      \hline
      \multicolumn{1}{|c|}{\ \ $\ell$ \ \ }& \multicolumn{1}{|c|}{\ \ $n$ \ \ }& \multicolumn{1}{|c|}{\ \ Third order WKB \ \ }&\multicolumn{1}{|c|}{\ \ Sixth order WKB \ \ }&\multicolumn{1}{|c|}{\ \ \ \ \ \ Prony \ \ \ \ \ \ } \\
      \hline
      $2$ & $0$ & $0.7346-0.0949i$ & $0.7347-0.0949i$&$ \ \ 0.7348-0.0949i \ \ $ \\
      \hline
  $2$ & $1$ & $0.7206-0.2870i$ & $0,7210-0.2869i$&$ 0,7210-0.2869i$ \\
      \hline
  $3$ & $0$ & $0.9343-0.0954i$ & $0.9344-0.0954i$&$ 0.9344-0.0953i$ \\
      \hline
 $3$ & $1$ & $0.9233-0.2876i$ & $0.9235-0.2876i$&$ 0.9237-0.2875i$ \\
     \hline
     $3$ & $2$ & $0.9031-0.4835i$ & $0.9026-0.4840i$&$ - $ \\
      \hline
 $4$ & $0$ & $1.1315-0.0956i$ & $1.1315-0.0956i$&$ 1.1317-0.0956i$ \\
      \hline
$4$ & $1$ & $1.224-0.2879i$ & $1.1225-0.2879i$&$ 1.1227-0.2878i$ \\
     \hline
       $4$ & $2$ & $1.1053-0.4828i$ & $1.1050-0.4831i$&$ - $ \\
      \hline
 $4$ & $3$ & $1.0817-0.6812i$ & $1.0798-0.6830i$&$-$ \\
      \hline
$5$ & $0$ & $1.3273-0.0958i$ & $1.3273-0.0958i$&$ 1.3276-0.0957i$ \\
     \hline
      $5$ & $1$ & $1.3196-0.2881i$ & $1.3196-0.2881i$&$ 1.3199-0.2879i$ \\
      \hline
       $5$ & $2$ & $1.3048-0.4824i$ & $1.3045-0.4826i$&$ - $ \\
      \hline
 $5$ & $3$ & $1.2839-0.6795i$ & $1.2826-0.6805i$&$ - $ \\
     \hline
     $5$ & $4$ & $1.2582-0.8794i$ & $1.2548-0.8832i$&$-$ \\
     \hline
       \hline
   \end{tabular}\label{frecuencias gravitino}}
      \end{table}
calculated the quasinormal frequencies of scalar, electromagnetic, gravitational and dirac perturbations in the Schwarzschild spacetime.

To comparative purposes, we also calculated the above frequencies using sixth order WKB method, and correct the results presented in reference \cite{konoplya1} for the quasinormal frequencies of Dirac and gravitino perturbations, where apparently the author uses an improper generalization of the formula for the effective potential due to boson fields in Schwarzshild spacetime to the case of fermion fields.
The equation for the radial part of boson perturbations have the form (\ref{finaleq}) with the effective potential:
\begin{equation}\label{bosonpot}
    V_{s}\left(r\right)=\frac{\triangle}{r^{4}}\left[\lambda+\beta\frac{r_{H}}{r}\right]
\end{equation}
where $\lambda=\ell(\ell+1)$ and the quantity $\beta$ is related to the spin-weight $s$ of the perturbing field by $\beta=1-s^{2}$. For scalar, electromagnetic and axial gravitational perturbations $\beta$ has the integer value 1, 0, -3, respectively.

\begin{table}[htb!]
    \tbl{\it Scalar quasinormal frequencies in Schwarzschild black hole, with
   $M=1$ (numerical and sixth order WKB approximation and Prony fitting of time domain data.}
  {\begin{tabular}{|c|c|c|c|c|}
      \hline
      \hline
      \multicolumn{1}{|c|}{\ \ $\ell$ \ \ }& \multicolumn{1}{|c|}{\ \ $n$ \ \ }& \multicolumn{1}{|c|}{\ \ Leaver results \ \ }&\multicolumn{1}{|c|}{\ \ Sixth order WKB \ \ }&\multicolumn{1}{|c|}{\ \ \ \ \ \ Prony \ \ \ \ \ \ } \\
      \hline
      $2$ & $0$ & $0.4836-0.0968i$ & $0.4836-0.0968i$&$ \ \ 0.4836-0.0968i \ \ $ \\
      \hline
  $2$ & $1$ & $0.4639-0.2956i$ & $0,4638-0.2956i$&$ 0.4639-0.2956i$ \\
      \hline
  $3$ & $0$ & $-$ & $0.6754-0.0965i$&$ 0.6754-0.0965i$ \\
      \hline
 $3$ & $1$ & $-$ & $0.6607-0.2923i$&$ 0.6609-0.2922i$ \\
     \hline
     $3$ & $2$ & $-$ & $0.6336-0.4960i$&$ - $ \\
      \hline
 $4$ & $0$ & $-$ & $0.8674-0.0964i$&$ 0.8674-0.0964i$ \\
      \hline
$4$ & $1$ & $-$ & $0.8558-0.2909i$&$ 0.8558-0.2909i$ \\
     \hline
       $4$ & $2$ & $-$ & $0.8337-0.4903i$&$ - $ \\
      \hline
 $4$ & $3$ & $-$ & $0.8032-0.6975i$&$-$ \\
      \hline
$5$ & $0$ & $-$ & $1.0596-0.0963i$&$ 1.0596-0.0963i$ \\
     \hline
      $5$ & $1$ & $-$ & $1.0500-0.2901i$&$ 1.0500-0.2901i$ \\
      \hline
       $5$ & $2$ & $-$ & $1.0315-0.4873i$&$ - $ \\
      \hline
 $5$ & $3$ & $-$ & $1.0052-0.6899i$&$ - $ \\
     \hline
     $5$ & $4$ & $-$ & $0.9728-0.8995i$&$-$ \\
     \hline
       \hline
   \end{tabular}\label{frequencias_escalar}}
     \end{table}

The polar perturbations corresponding to a gravitational perturbing field are governed by the so called Zerilli potential, that is different from (\ref{bosonpot}), but as was rigorously shown by Chandrasekar in \cite{chandrasekar1}, the quasinormal frequencies belonging to axial and polar perturbations are identical.

For the case of Dirac perturbing field, we use the form of the effective potential obtained by Cho in reference \cite{cho}, given by:
\begin{equation}\label{diracpot1}
    V_{\frac{1}{2}}\left(r\right)=\frac{dW}{dr_{*}}+W^{2}
\end{equation}
with
\begin{equation}\label{diracpot2}
    W\left(r\right)=\kappa\frac{\sqrt{\triangle}}{r^{4}}
\end{equation}
where the quantity $\kappa=\ell+1$.
Tables \ref{frequencias_escalar} to \ref{frequencias_gravitacionales} shows the results obtained. We also show the third order WKB results for all the perturbations. In the case of scalar, electromagnetic and gravitational perturbations, we listed the numerical results given in the paper of Leaver \cite{leaver1}.

The third order WKB results showed for the case of Dirac perturbations are a refinement of the Cho results \cite{cho}, to include one more decimal place (see also \cite{shu}).
\begin{figure}[htb!]
\begin{center}
\resizebox{0.49\columnwidth}{!}{\includegraphics*{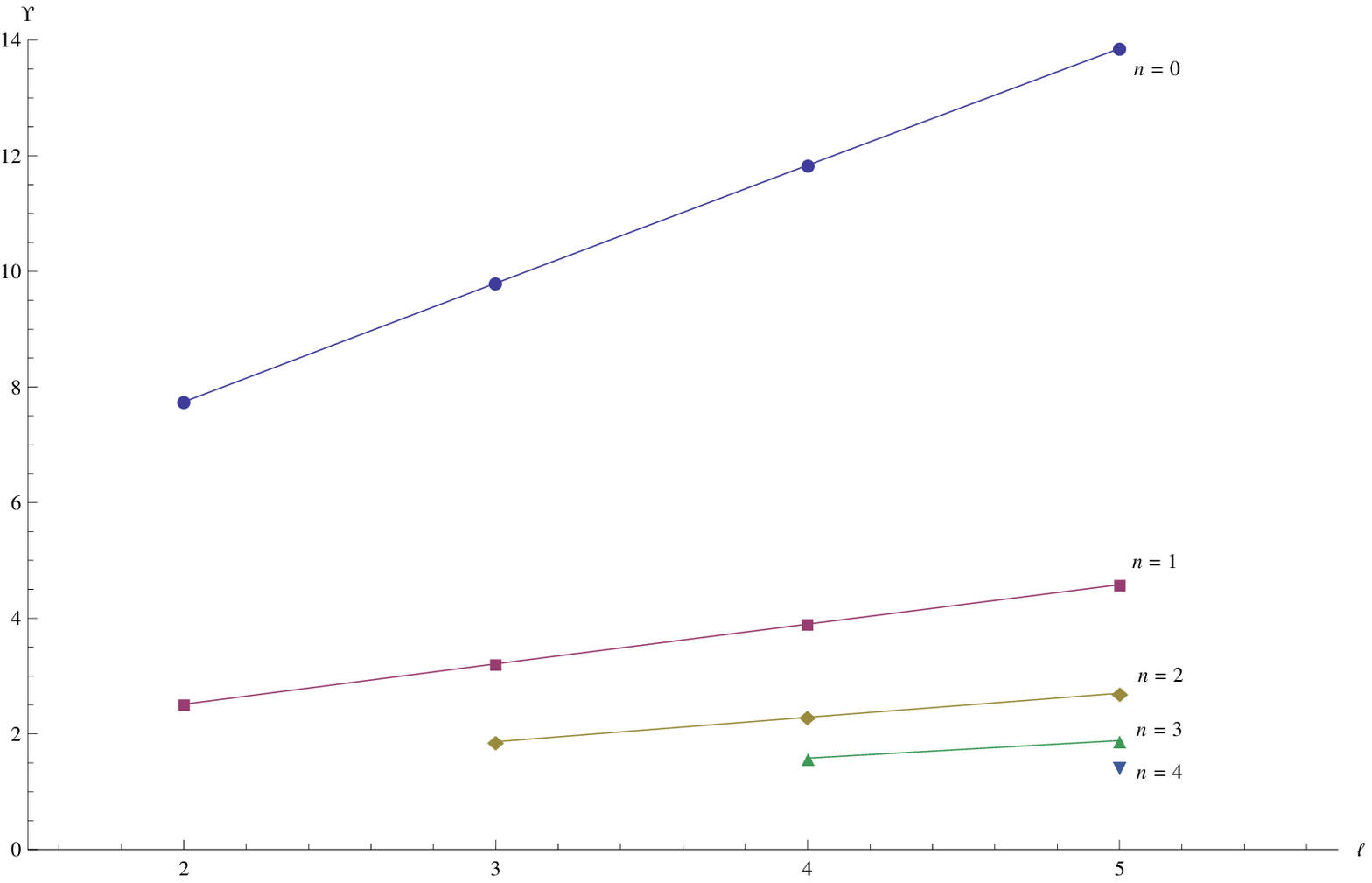}}
\resizebox{0.49\columnwidth}{!}{\includegraphics*{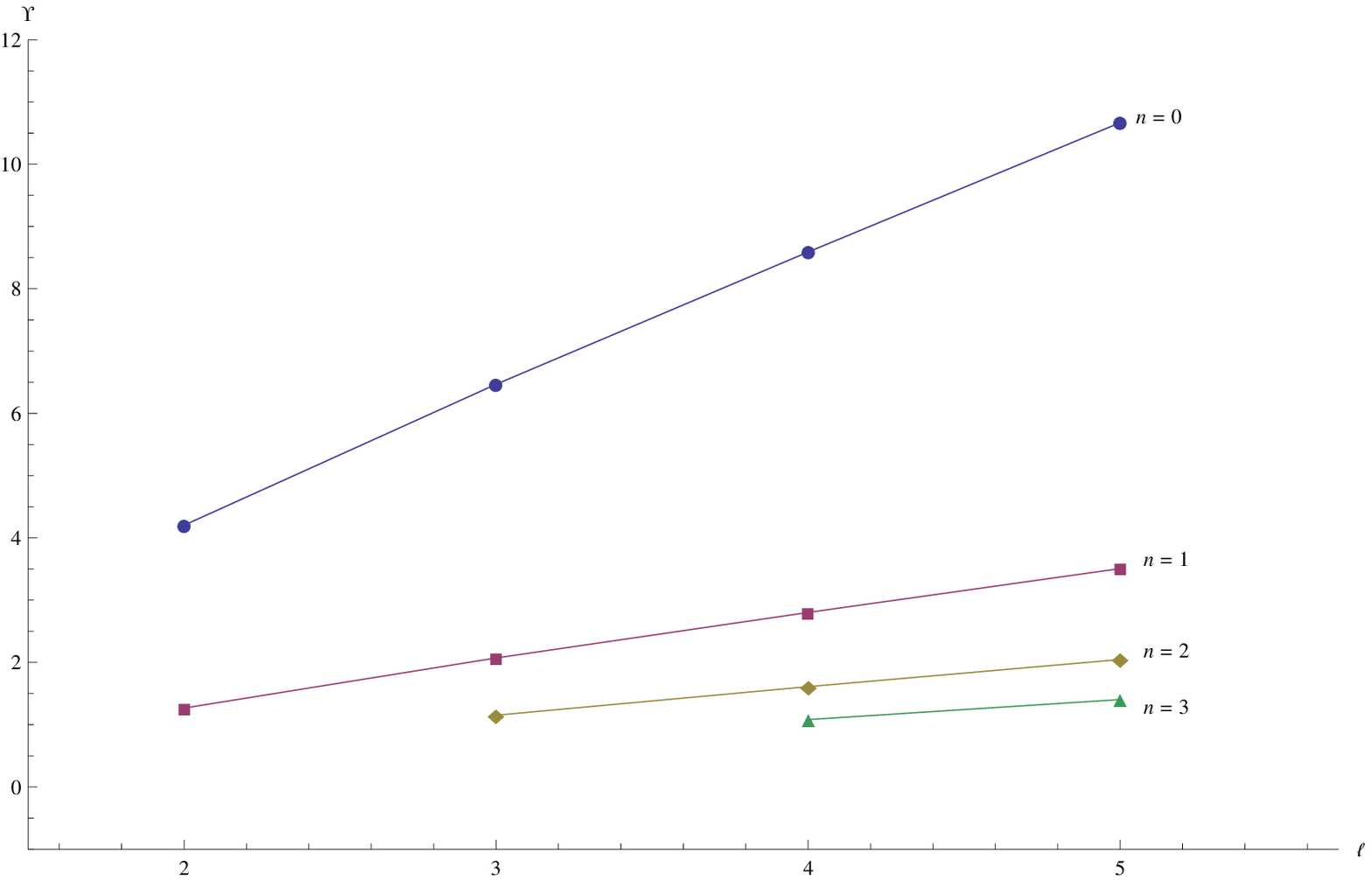}}
\caption{Dependance of the quality factor upon multipole and overtone numbers for gravitino (left) and gravitational (right) perturbations.}
\label{qfactor}
\end{center}
\end{figure}
\begin{figure}[htb!]
\begin{center}
\resizebox{0.49\columnwidth}{!}{\includegraphics*{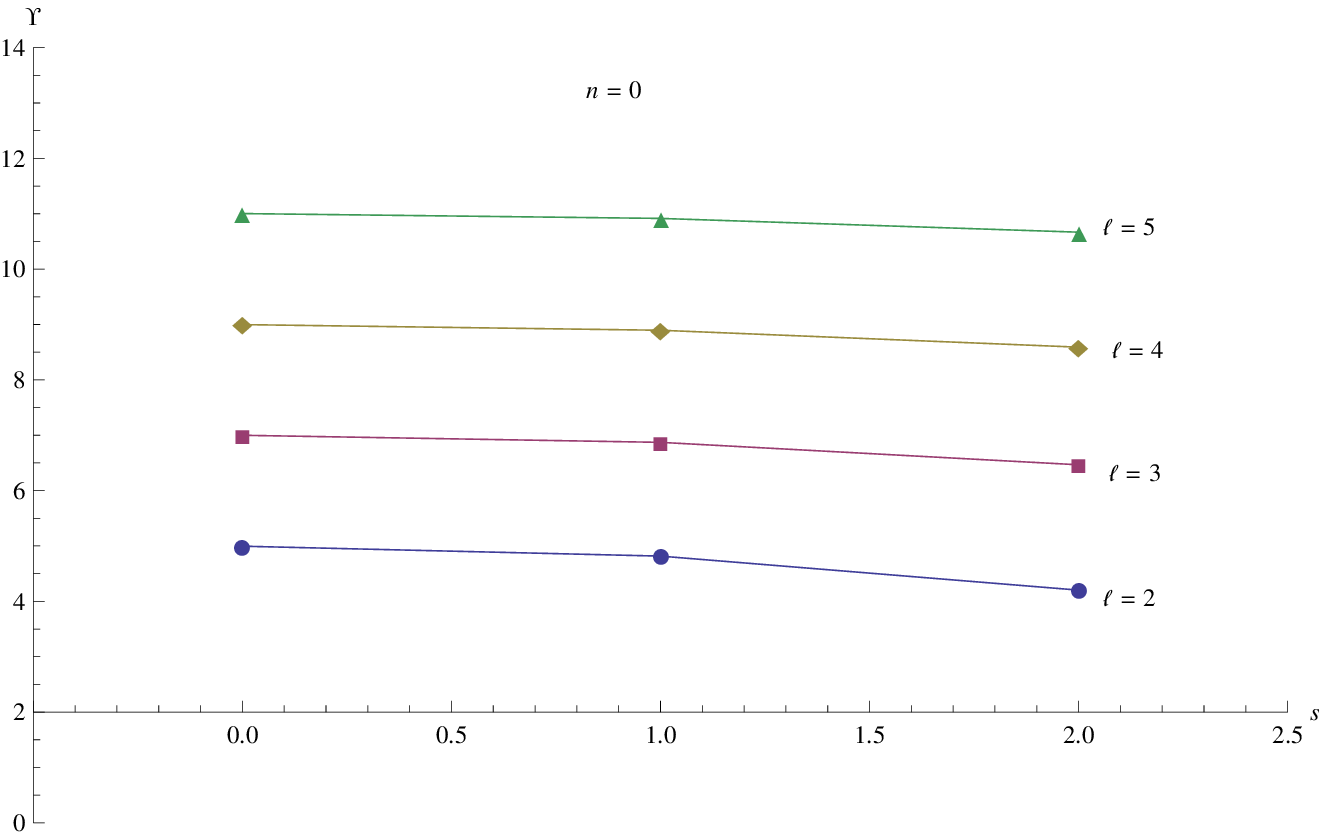}}
\resizebox{0.49\columnwidth}{!}{\includegraphics*{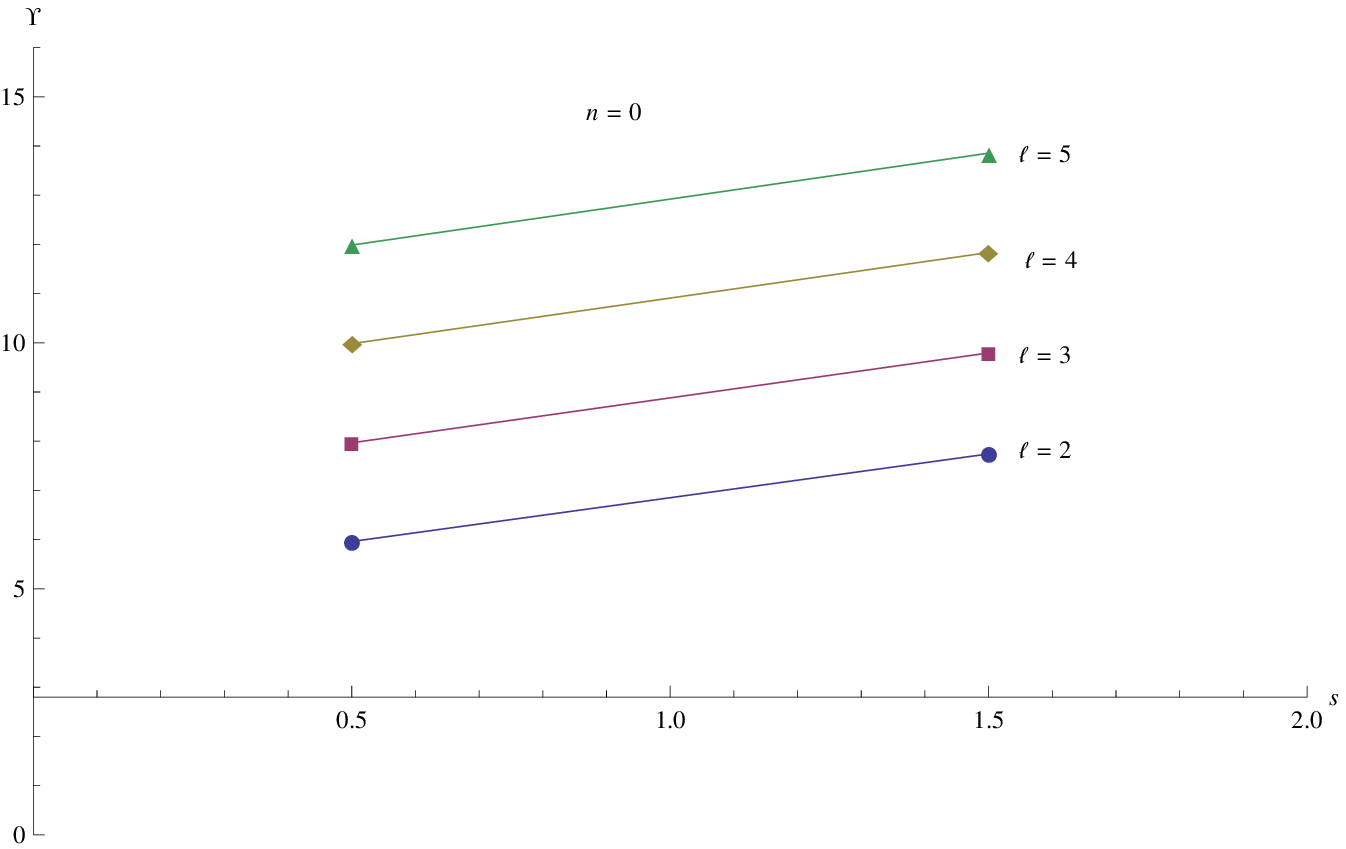}}
\caption{Dependance of the quality factor upon spin for boson(left) and fermion(right) perturbations.}
\label{qfactorspin}
\end{center}
\end{figure}
\begin{table}[htb!]
    \tbl{\it Electromagnetic quasinormal frequencies in Schwarzschild black hole, with
   $M=1$ (numerical and sixth order WKB approximation and Prony fitting of time domain data).}
  {\begin{tabular}{|c|c|c|c|c|}
      \hline
      \hline
      \multicolumn{1}{|c|}{\ \ $\ell$ \ \ }& \multicolumn{1}{|c|}{\ \ $n$ \ \ }& \multicolumn{1}{|c|}{\ \ Leaver results \ \ }&\multicolumn{1}{|c|}{\ \ Sixth order WKB \ \ }&\multicolumn{1}{|c|}{\ \ \ \ \ \ Prony \ \ \ \ \ \ } \\
      \hline
      $2$ & $0$ & $0.4576-0.0950i$ & $0.4576-0.0950i$&$ \ \ 0.4576-0.0950i \ \ $ \\
      \hline
  $2$ & $1$ & $0.4365-0.2907i$ & $0,4365-0.2907i$&$ 0.4365-0.2907i$ \\
      \hline
  $3$ & $0$ & $0.6569-0.0956i$ & $0.6569-0.0956i$&$ 0.6569-0.0956i$ \\
      \hline
 $3$ & $1$ & $0.6417-0.2897i$ & $0.6417-0.2897i$&$ 0.6417-0.2897i$ \\
     \hline
     $3$ & $2$ & $0.6138-0.4921i$ & $0.6138-0.4921i$&$ - $ \\
      \hline
 $4$ & $0$ & $-$ & $0.8531-0.0959i$&$ 0.8531-0.0959i$ \\
      \hline
$4$ & $1$ & $-$ & $0.8413-0.2893i$&$ 0.8414-0.2893i$ \\
     \hline
       $4$ & $2$ & $-$ & $0.8187-0.4878i$&$ - $ \\
      \hline
 $4$ & $3$ & $-$ & $0.7876-0.6942i$&$-$ \\
      \hline
$5$ & $0$ & $-$ & $1.0479-0.0960i$&$ 1.0479-0.0959i$ \\
     \hline
      $5$ & $1$ & $-$ & $1.0382-0.2891i$&$ 1.0384-0.2890i$ \\
      \hline
       $5$ & $2$ & $-$ & $1.0194-0.4856i$&$ - $ \\
      \hline
 $5$ & $3$ & $-$ & $0.9928-0.6876i$&$ - $ \\
     \hline
     $5$ & $4$ & $-$ & $0.9510-0.8967i$&$-$ \\
     \hline
       \hline
   \end{tabular}\label{frequencias_electromagnetico}}
     \end{table}
\begin{table}[htb!]
    \tbl{\it Dirac quasinormal frequencies in Schwarzschild black hole, with
   $M=1$ (numerical and sixth order WKB approximation and Prony fitting of time domain data).}
  {\begin{tabular}{|c|c|c|c|c|}
      \hline
      \hline
      \multicolumn{1}{|c|}{\ \ $\ell$ \ \ }& \multicolumn{1}{|c|}{\ \ $n$ \ \ }& \multicolumn{1}{|c|}{\ \ Third order WKB \ \ }&\multicolumn{1}{|c|}{\ \ Sixth order WKB \ \ }&\multicolumn{1}{|c|}{\ \ \ \ \ \ Prony \ \ \ \ \ \ } \\
      \hline
      $2$ & $0$ & $0.5737-0.0963i$ & $0.5741-0.0963i$&$ \ \ 0.5743-0.0962i \ \ $ \\
      \hline
  $2$ & $1$ & $0.5562-0.2930i$ & $0.5570-0.2927i$&$ 0.5571-0.2927i$ \\
      \hline
  $3$ & $0$ & $0.7672-0.0963i$ & $0.7673-0.0963i$&$ 0.7675-0.0962i$ \\
      \hline
 $3$ & $1$ & $0.7540-0.2910i$ & $0.7543-0.2910i$&$ 0.7543-0.2910i$ \\
     \hline
     $3$ & $2$ & $0.7304-0.4909i$ & $0.7297-0.4919i$&$ - $ \\
      \hline
 $4$ & $0$ & $0.9602-0.0963i$ & $0.9603-0.0962i$&$ 0.9605-0.0961i$ \\
      \hline
$4$ & $1$ & $0.9496-0.2902i$ & $0.9498-0.2901i$&$ 0.9499-0.2901i$ \\
     \hline
       $4$ & $2$ & $0.9300-0.4876i$ & $0.9295-0.4881i$&$ - $ \\
      \hline
 $4$ & $3$ & $0.9036-0.6892i$ & $0.9011-0.6925i$&$-$ \\
      \hline
$5$ & $0$ & $-$ & $1.1531-0.0962i$&$ 1.1531-0.0962i$ \\
     \hline
      $5$ & $1$ & $-$ & $1.1442-0.2897i$&$ 1.1443-0.2897i$ \\
      \hline
       $5$ & $2$ & $-$ & $1.1271-0.4860i$&$ - $ \\
      \hline
 $5$ & $3$ & $-$ & $1.1025-0.6869i$&$ - $ \\
     \hline
     $5$ & $4$ & $-$ & $1.0718-0.8939i$&$-$ \\
     \hline
       \hline
   \end{tabular} \label{frequencias_dirac}}
     \end{table}

\begin{table}[htb!]
    \tbl{\it Gravitational quasinormal frequencies in Schwarzschild black hole, with
   $M=1$ (numerical and sixth order WKB approximation and Prony fitting of time domain data).}
  {\begin{tabular}{|c|c|c|c|c|}
      \hline
      \hline
      \multicolumn{1}{|c|}{\ \ $\ell$ \ \ }& \multicolumn{1}{|c|}{\ \ $n$ \ \ }& \multicolumn{1}{|c|}{\ \ Leaver results \ \ }&\multicolumn{1}{|c|}{\ \ Sixth order WKB \ \ }&\multicolumn{1}{|c|}{\ \ \ \ \ \ Prony \ \ \ \ \ \ } \\
      \hline
      $2$ & $0$ & $0.3737-0.0890i$ & $0.3736-0.0889i$&$ \ \ 0.3737-0.0890i \ \ $ \\
      \hline
  $2$ & $1$ & $0.3467-0.2739i$ & $0,3463-0.2735i$&$ 0.3467-0.2739i$ \\
      \hline
  $3$ & $0$ & $0.5994-0.0927i$ & $0.5994-0.0927i$&$ 0.5994-0.0927i$ \\
      \hline
 $3$ & $1$ & $0.5826-0.2813i$ & $0.5826-0.2813i$&$ 0.5826-0.2813i$ \\
     \hline
     $3$ & $2$ & $0.5517-0.4791i$ & $0.5516-0.4790i$&$ - $ \\
      \hline
 $4$ & $0$ & $0.8092-0.0942i$ & $0.8092-0.0942i$&$ 0.8092-0.0942i$ \\
      \hline
$4$ & $1$ & $0.7966-0.2843i$ & $0.7966-0.2843i$&$ 0.7966-0.2843i$ \\
     \hline
       $4$ & $2$ & $0.7727-0.4799i$ & $0.7727-0.4799i$&$ - $ \\
      \hline
 $4$ & $3$ & $0.7398-0.6839i$ & $0.7397-0.6839i$&$-$ \\
      \hline
$5$ & $0$ & $-$ & $1.0123-0.0949i$&$ 1.0123-0.0949i$ \\
     \hline
      $5$ & $1$ & $-$ & $1.0022-0.2858i$&$ 1.0022-0.2858i$ \\
      \hline
       $5$ & $2$ & $-$ & $0.9827-0.4803i$&$ - $ \\
      \hline
 $5$ & $3$ & $-$ & $0.9550-0.6805i$&$ - $ \\
     \hline
     $5$ & $4$ & $-$ & $0.9208-0.8881i$&$-$ \\
     \hline
       \hline
   \end{tabular}\label{frequencias_gravitacionales}}
     \end{table}

As we can see, the Prony method gives values that are in perfect agreement with previous numerical results existing for scalar, electromagnetic and gravitational perturbations. Also we obtain a improvement of the third order WKB results using the sixth order semianalytic formula.

The dependance of the quality factor with the multipole and overtone number is similar as for the case of gravitino test field perturbations. In Figure (\ref{qfactor}) we show this dependance for gravitino and gravitational perturbations. For a fixed multipole number, higher overtones shows a decreasing in the quality factor, and the modes are more damping. In contrast, increasing the multipole number for a given overtone, the quality factor increases due to the increasing of the real oscillating frequency in a more pronounced way that the damping rate.
At this point it is interesting to note that fermion perturbations in the black hole background shows higher quality factors that boson perturbations, as Figure (\ref{qfactorspin}) shows.
As the spin weight of a boson perturbation increases, the quality factor decreases, and the black holes perturbed by this fields becomes more poor oscillators. Then, the lowest quality factors corresponds to gravitational perturbations, and the higher to scalar test perturbations.

The situation for fermion fields is opposite. In this case the quality factor increases with the increment of the field spin, and as a consequence the highest quality factor belongs to the gravitino perturbations. Then, as a general fact, black holes perturbed by fermion fields are better oscillators than those perturbed by boson fields.

\section{Concluding remarks}
We have studied the evolution of Rarita-Schwinger field
perturbations in a Schwarzschild background. Solving numerically the
time evolution equation for this perturbations, we find similar time
domain profiles as in the case of fields of other spins: the usual
three stages in the time evolution dominated at intermediary times
by quasinormal ringing. We determined the quasinormal frequencies by
two different approaches, 6th order WKB and time domain integration
with Prony fitting of the numerical data. The two methods give close values of QNMs for well pronounced potential
barriers.

Also we apply the Prony method to numerically calculated the quasinormal frequencies of scalar, electromagnetic, gravitational and Dirac perturbations in the Schwarzschild background, with results in perfect agreement with previous numerical calculations by other authors.

The time domain evolution of gravitino perturbations are similar to that corresponding to other fields of different spin. The behaviour of the quasinormal frequencies with respect to the black holes masses is also similar for all spin perturbations. However, an interesting difference occur in the oscillating behaviour of black holes perturbed by fields of different spin classes: if the perturbing field has integer spin, then the system becomes a poor oscillator as the spin increases. The opposite situation occurs in the presence of fermion perturbations, black holes perturbed by higher spin fields are better oscillators. In general, the highest quality factor is associated with gravitino perturbations and the smallest with gravitational ones.

There are extensions of this work that are interesting to
consider, first the generalization to the case of charged black holes, and second, the determination of the changes of the gravitino
quasinormal spectrum for semiclassical solutions, due to the vacuum polarization of quantized fields. The solution of the above
problems will be presented in future reports.

\section{Acknowledgements}
We are grateful to Dr. Alexander Zhidenko from USP, Brazil for
providing me with his MATHEMATICA code with the implementation of
the Prony method and valuable information about numerical methods
usually employed in the calculation of quasinormal frequencies. Also we would like to acknowledge helpful discussions with Elcio Abdalla and Jeferson de Oliveira from Department of Mathematical-Physics at the University of S\~{a}o Paulo, where this work was completed .


\end{document}